\documentclass{jfm}

\usepackage{graphicx,
natbib,
upmath,
bm,
amsmath,
amssymb,
amsbsy,
color}
\usepackage[makeroom]{cancel}

\newcommand\NavSto{Navier--Stokes}

\newcommand\TG{Taylor--G\"{o}rtler}

\newcommand\Rey{\mbox{\textit{Re}}}

\newcommand{\eg}{e.g.\ }
\newcommand{\CC}{\mathrm{c.c.}}

\newcommand\ce{\mathrm{e}}
\newcommand\ci{\mathrm{i}}

\newcommand\twod{two-di\-men\-sion\-al}
\newcommand\threed{three-di\-men\-sion\-al}

\title[A reduced-order model based on the resolvent operator]{A
  reduced-order model of three-dimensional unsteady flow in a cavity
  based on the resolvent operator}

\author[F. G\'omez and others]{F. G\'omez$^1$\thanks{Email address for
correspondence: francisco.gomez-carrasco@monash.edu},
  H. M. Blackburn$^{1}$, M. Rudman$^1$,\break A. S. Sharma$^2$
  and B. J. McKeon$^3$}

\affiliation{$^1$Dept. of Mechanical and Aerospace Engineering, Monash
  University, VIC 3800, Australia\\[\affilskip] $^2$Faculty of
  Engineering and the Environment, \\ University of Southampton,
  Southampton SO17 1BJ, UK \\[\affilskip] $^3$Graduate Aerospace
  Laboratories, \\ California Institute of Technology, Pasadena, CA
  91125, USA }

\date{?; revised ?; accepted ?. - To be entered by editorial office}

\begin{document}

\maketitle

\begin{abstract}
A novel reduced-order model for time-varying nonlinear flows arising
from a resolvent decomposition based on the time-mean flow is
proposed.  The inputs required for the model are the mean flow field
and a small set of velocity time-series data obtained at isolated
measurement points, which are used to fix relevant frequencies,
amplitudes and phases of a limited number of resolvent modes that,
together with the mean flow, constitute the reduced-order model.  The
technique is applied to derive a model for the unsteady \threed\ flow
in a lid-driven cavity at a Reynolds number of 1200 that is based on
the \twod\ mean flow, three resolvent modes selected at the most
active spanwise wavenumber, and either one or two velocity probe
signals.  The least-squares full-field error of the reconstructed
velocity obtained using the model and two point velocity probes is of
order 5\% of the lid velocity, and the dynamical behaviour of the
reconstructed flow is qualitatively similar to that of the complete
flow.\\[6pt]
\end{abstract}

\section{Introduction}
\label{sec.intro}

The development of reduced-order models (ROMs) to represent physics of
fluid flows is a subject of considerable current interest in fluid
mechanics. The construction of such models is motivated \eg by
potential application to flow control for drag reduction and noise
suppression \citep{brunton2015closed}. Spatial shapes that serve as
bases for ROMs can be classified as mathematical, empirical or
physical modes. Mathematical modes form a complete basis by definition
and many ROMs based on expansion functions have been used for simple
boundary conditions \citep{busse1991numerical,noack1994low}.
Empirical ROMs such as the proper orthogonal decomposition
\citep*[POD,][]{berkooz1993proper} or the dynamic mode decomposition
\citep[DMD,][]{Schmid2010} arise from post-processing of numerical or
experimental flow data. The present work expounds a novel ROM based on
physical modes emergent from the \NavSto\ equations.

\citet*{noack2011reduced} proposed that linear global stability
analysis could be employed to obtain physical modes associated with
linear dynamics from the \NavSto\ equations.
Such analyses are based on a decomposition of the flow into a steady
or periodic laminar base flow and an infinitesimal perturbation that
develops in time, leading to an eigenvalue problem whose eigenvalues
characterize the stability of the base flow.  The eigenmodes can be
employed as spatial shapes to construct ROMs.
While computationally demanding if the base flow is \threed, spatially
complicated, or the Reynolds number is large, methodologies for
numerical linear global stability analyses are becoming mature.

A different type of challenge arises from the application of global
stability analysis to turbulent, or more generally, to nonlinear
flows. By nonlinear flows, we mean unsteady flows in which there exist
different frequencies that interact with each other; feedback via the
nonlinear terms in the \NavSto\ equations is relevant. The key step
enabling global stability analysis of nonlinear flows is to consider
the time-mean flow as the base flow.
\citet{barkley2006linear} applied global stability analysis to the
wake of a circular cylinder to study the shedding frequency at
Reynolds numbers above the onset of vortex shedding. It was observed
that linear stability analysis of unstable (symmetric) steady base
flows provided frequencies and global modes different to those
observed in numerical simulations and experiments.
However, when applied to the applied to the mean flow, the method was
able to predict frequencies and flow structures similar to those
observed in numerical simulations and experiments.
More recently, \citet*{oberleithner2014mean} applied the same
methodology to jets.
Despite apparent successes of the method in these cases, global
stability analysis applied to mean flows can be considered
dubious for two reasons.

The first reason is that time-averaged flows are not solutions of the
\NavSto\ equations, but of the Reynolds-averaged \NavSto\ (RANS)
equations. The closure problem is thus inherent in the approach and
the unknown Reynolds stress, arising from the interaction of turbulent
fluctuations with the mean, must be considered. One way to avoid this
issue for weakly nonlinear flows is to employ the assumption, proposed
by \citet{barkley2006linear}, that the forcing generated by the
fluctuations is steady; it only contributes as Reynolds stress in the
mean flow equation. Another approach is to close the RANS equations
via the Boussinesq hypothesis and employ an Newtonian eddy viscosity
to account for the Reynolds stress.  However, this only applies for
fully developed turbulent flows in which the diffusion induced by the
turbulence can be approximated by an additional eddy viscosity. 
This approach was introduced in conjunction with a triple
decomposition by \citet{hussain1970mechanics} in order to identify
coherent flow structures in turbulent shear flows 
and more recently employed \eg by \citet*{meliga2012sensitivity} in an
flow control context.
To our knowledge, no existing method is able to account for
nonlinearity in flows which are neither weakly nonlinear nor fully
turbulent.

The second concern with mean flow stability analysis is with its
reliability. \citet{sipp2007global} replicated
\citeauthor{barkley2006linear}'s work on the cylinder mean flow, then
applied the same methodology to an open cavity flow.  In the latter
case, the predicted frequencies did not match those observed in direct
numerical simulation (DNS) and attributed the discrepancy to the
relative strength of the mean flow and harmonics, linked to the
non-normality of the flow, such that the global modes obtained are
non-orthogonal. This is not a desirable characteristic when choosing a
ROM since the projection of flow solutions onto global modes would be
ill-conditioned, as discussed by \citet{cerqueira2014eigenvalue}.

In the following we show that the resolvent analysis for turbulent
flows developed by \citet{McKeonSharma2010} represents an alternative
strategy to identify physical flow structures in nonlinear flows. The
methodology consists of an amplification analysis of the
\NavSto\ equations in the frequency domain, yielding a linear
relationship between the velocity field at specific wavenumbers and
the nonlinear interaction between other wavenumbers via a resolvent
operator.  A singular value decomposition (SVD) of the operator
reveals that it acts as a very selective directional amplifier of the
nonlinear terms and hence a low-rank approximation of the resolvent is
able to reproduce the dominant features of the flow. The time-mean
flow must be given as input.

Resolvent analyses have been successfully employed to qualitatively
describe the behavior of flow structures in turbulent canonical flows
\citep{SharmaMcKeon2013} and to identify sparsity effects in direct
numerical simulation with periodic finite-length domain
\citep{gomez_pof_2014}. The technique has also been applied to model
scalings in turbulent flows \citep{Moarref2013} and to wall-based
closed-loop flow control strategies
\citep*{luhar2014opposition,luhar2015framework}.  We will explain how
the analysis circumvents the above-mentioned limitations of the
mean-flow global stability analysis in a ROM context. As an example,
we demonstrate the potential of this methodology by employing a
\twod\ resolvent formulation
\citep{aakervik2008global,brandt2011effect,gomez_pof_2014} to obtain
the relevant flow features in a \threed\ spanwise-periodic lid-driven
cavity flow. For this purpose, a ROM is derived by employing the mean
flow and minimal temporal and spatial spectral information.

Lid-driven cavity flows can possess complex features despite the
simple geometry, hence they serve as a model problem for many
engineering applications dealing with flow recirculation. From the
work of \citet{KoseffStreet84b}, it is known that most of the velocity
fluctuation in the lid-driven cavity is caused by \TG-like (TGL)
vortices. The naming of these vortices is justified by resemblances of
the \twod\ base flow and the corresponding \threed\ features with the
respective flows in the Taylor and G\"{o}rtler problems, as discussed by
\citet{albensoeder2006nonlinear}
in their study of the non-linear stability boundaries of the
flow. Depending on the spanwise length and Reynolds number, different
numbers of TGL pairs of vortices can co-exist in the flow.  In the
following, we select flow conditions such that the flow possesses
structures with distinct associated frequencies that interact, but
which is far from a turbulent state.  Resolvent analysis is be
employed to construct a reduced-order model of this flow.

\section{Description of the flow}
\label{sec.flow}

\begin{figure}
\begin{center} 
\includegraphics[width=0.8\linewidth]{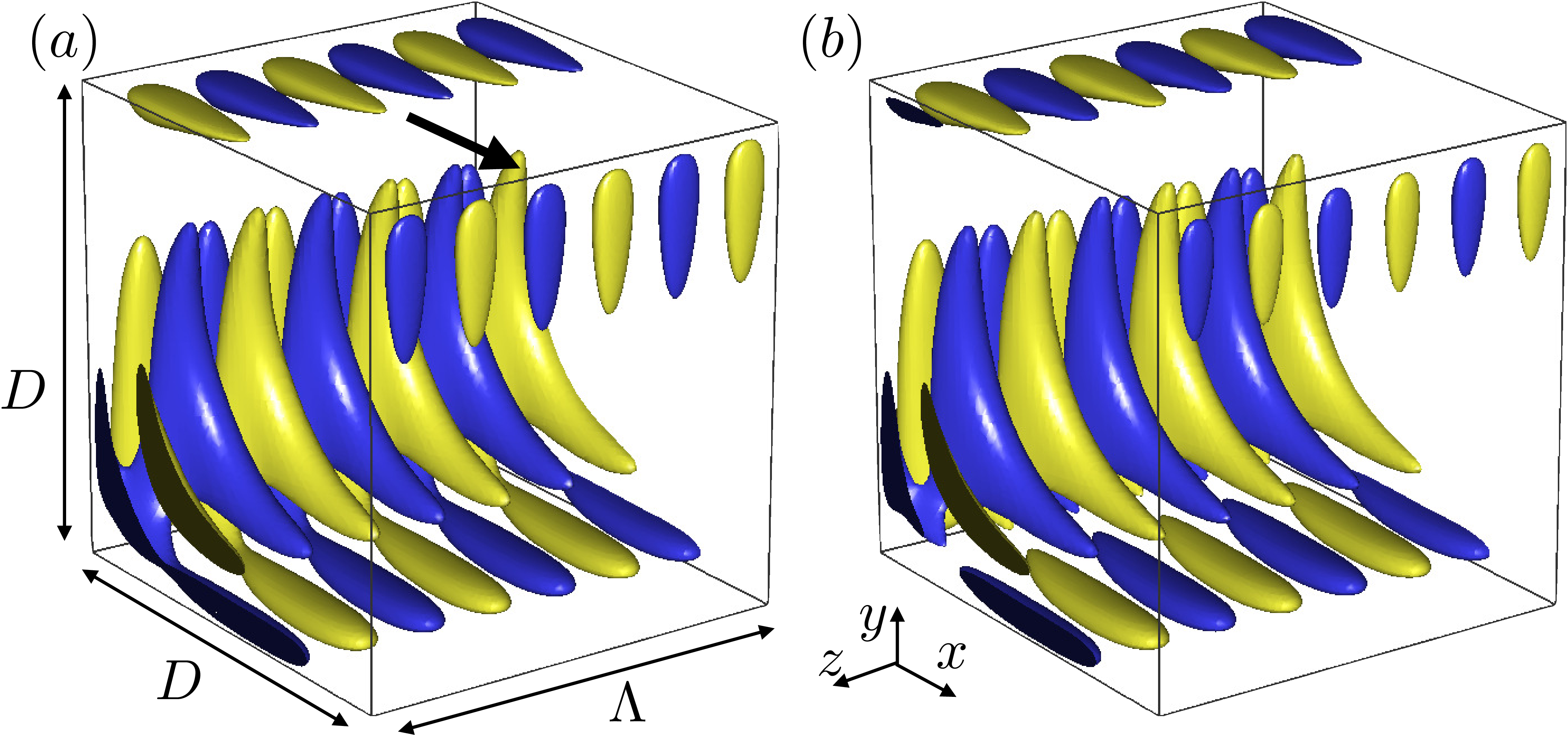} 
\end{center}
\caption{Representation of TGL vortices in the lid-driven cavity flow
  at $\Rey=1200$ and $\Lambda/D=0.945$; lid moves in the direction of
  arrow.  Snapshot of iso-surfaces of $20\%$ max/min spanwise velocity
  of spanwise Fourier mode $\beta=3$. (a) DNS, (b) resolvent-based
  model. Animations of iso-surfaces of DNS and model are shown in
  supplementary movies~1 and~2 respectively. }
\label{fig:LDC}
\end{figure}

The spanwise-periodic lid-driven cavity flow with a square
cross-section is governed by the \threed\ incompressible
\NavSto\ equations
\begin{equation}
  \partial_t\bm{\hat{u}}+\bm{\hat{u}\cdot\nabla\hat{u}} =
       -\bm{\nabla} {\hat{p}} +
       \Rey^{-1}\nabla^{2}{\bm{\hat{u}}},\qquad
  \bm{\nabla\cdot u}=0, \label{eqn:NSE}
\end{equation} 
where $\Rey$ is the Reynolds number based on the steady lid speed $U$
and cavity depth $D$, $\bm{\hat{u}}=(u,v,w)$ is the velocity vector
expressed in Cartesian coordinates $(x,y,z)$ and $\hat{p}$ is the
modified pressure. The geometry is illustrated in
figure~\ref{fig:LDC}(\textit{a}), in which $\Lambda/D$ denotes the
periodic span of the cavity. No-slip boundary conditions are imposed at
the walls.

DNS of the incompressible lid-driven cavity flow has been carried out
using a spectral element--{Fourier} solver \citep{hugh2002}.  Nodal
elemental basis functions are used in the $(x,y)$ plane while a
Fourier expansion basis is employed in the homogeneous direction
$(z)$. The flow solution can be written as a sum of spanwise Fourier
modes
\begin{equation}
\bm{\hat{u}}(x,y,z,t)= 
\sum_{\beta} \bm{u}_{\beta}(x,y,t) \ce^{\ci\beta z} + \CC
\label{eq:2.5D}
\end{equation}
with $\beta$ a non-dimensional spanwise wavenumber normalized with the
span $\Lambda$. Based on the nonlinear stability boundaries
investigated by \citet{albensoeder2006nonlinear}, we chose parameters
$\Rey=1200$ and $\Lambda/D=0.945$ for our simulations. This selection
provides an unsteady flow with multiple frequencies and three pairs of
TGL vortices. Figure~\ref{fig:LDC}(\textit{a}) represents the spanwise
velocity isosurfaces of the flow and three pairs of vortices may be
identified. The temporal behavior of these vortices is shown in the
supplementary movie~1, which presents animations of spanwise velocity
isosurfaces.

\begin{figure}
\begin{center} 
\includegraphics[width=0.8\linewidth]{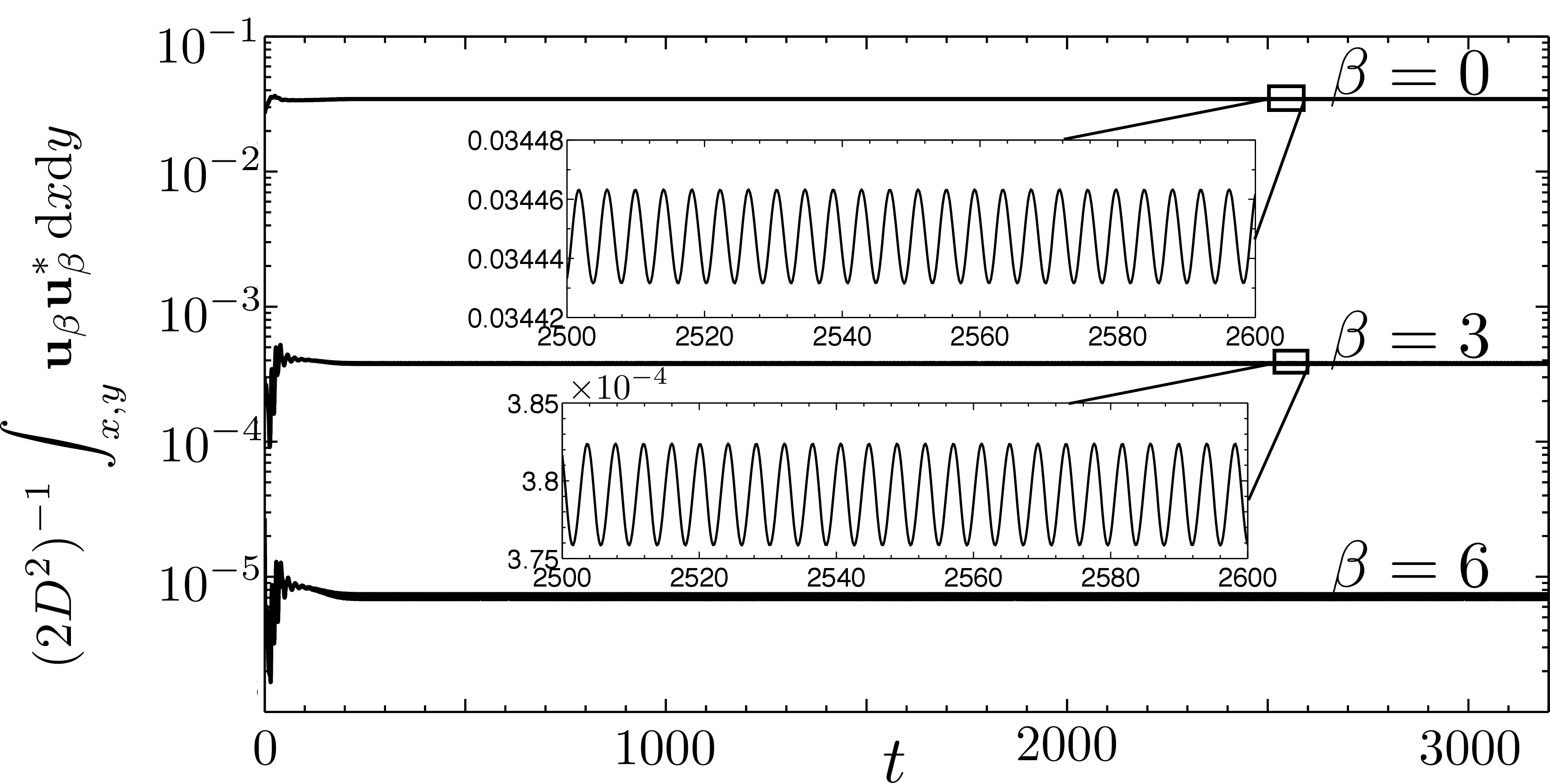} 
\end{center}
\caption{Temporal evolution of the kinetic energy based on
  (\ref{eq:2.5D}) of the three most energetic spanwise Fourier
  modes. Mode $\beta=0$ contains the mean flow while $\beta=3$ consist
  of three pairs of TGL vortices. Self-interaction of mode $\beta=3$
  provides energy to its harmonic $\beta=6$.  }
\label{fig:mdl}
\end{figure}

Figure \ref{fig:mdl} presents the temporal evolution of the kinetic
energy based on (\ref{eq:2.5D}) of the energetically relevant spanwise
Fourier modes. This measure indicates when the flow reaches a
statistically steady periodic state. Data for $t<300$ have been
discarded and flow statistics collected until those for the mean flow
converged. This \twod\ flow is the basis of our subsequent resolvent
analysis. Mode $\beta=0$ contains the mean flow while $\beta=3$
consist of three pairs of TGL vortices. Self-interaction of mode
$\beta=3$ provides energy to its harmonics, but the energy associated
with these modes is an order of magnitude smaller.

Spatial resolution convergence is achieved with a mesh consists of 132
spectral elements with polynomial degree $N_p=9$ in each the $(x,y)$
planes and 64 spanwise Fourier modes. 500 time steps are employed to
integrate a time unit $D/U$.

\section{Description of the model}
\label{sec.model}

\subsection{Resolvent analysis}

Here we extend the resolvent analysis of \citet{McKeonSharma2010} for
pipe flows to the spanwise-periodic \threed\ lid-driven cavity flow
under consideration. The flow is assumed to be statistically steady. A
spanwise and time averaged mean flow $\bm{u}_{0,0}$ is obtained from
the DNS and subtracted from the total velocity to leave the
fluctuating velocity $\bm{u}=\bm{\hat{u}}-\bm{u}_{0,0}$, which may be
decomposed as a sum of Fourier mode in spanwise direction and time
\begin{equation}
\bm{u}(x,y,z,t)= \sum_{\beta} \sum_{\omega} \bm{u}_{\beta,\omega}(x,y)
\ce^{\ci(\beta z -\omega t)} + \CC \, ,
\end{equation}
Both $\beta$ and $\omega$ are real and the continuous integral in
frequency has been expressed as a sum for simplicity. A similar
decomposition may be applied to the nonlinear terms, leading to
$\bm{f}_{\beta,\omega}=(\bm{u\cdot\nabla u})_{\beta,\omega}$.
The introduction of these decompositions into the \NavSto\
equations (\ref{eqn:NSE}) leads to
\begin{eqnarray}
0 &=& \bm{f}_{0,0} - \bm{u}_{0,0}\bm{\cdot\nabla u}_{0,0} +
\Rey^{-1}\nabla^2\bm{u}_{0,0} \label{mean} \, , \\ 
\ci \omega \bm{u}_{\beta,\omega} &=& 
\mathcal{L}_{\beta,\omega}\bm{ u}_{\beta,\omega} + \bm{f}_{\beta,\omega} \, ,
\label{LNSE+f}
\end{eqnarray}
with $\mathcal{L}_{\beta,\omega}$ being the Jacobian operator of the
\NavSto\ for each set of $(\beta,\omega)$. The equation corresponding
to $(\beta,\omega)=(0,0)$ is known as the RANS equation and the
Reynolds stress $\bm{f}_{0,0}$ denotes the interaction of the
fluctuating velocity with the mean. For clarity, the velocity is
projected onto a divergence-free basis in order to eliminate the
pressure.  Further re-arrangement of (\ref{LNSE+f}) reads
\begin{equation}
\bm{u}_{\beta,\omega} = \mathcal{H}_{\beta,\omega}\bm{f}_{\beta,\omega} \, ,
\label{eq:reseq}
\end{equation}
in which $\mathcal{H}_{\beta,\omega}= (\ci\,\omega -
\mathcal{L}_{\beta,\omega})^{-1}$ is known as the resolvent operator.
\begin{equation}
 \mathcal{{H}}_{\beta,\omega}(x,y) =\left[
\begin{array}{cccc}
D - \partial_x u_0 + \ci \omega  &  \partial_y{u_0} & 0 & -\partial_x \\
  \partial_x v_0 &  D - \partial_y v_0 + \ci \omega   & 0 & -\partial_y \\
  0 & 0 & D + \ci \omega & -\ci \beta  \\
\partial_x & \partial_y & \ci \beta & 0 
\end{array}  \right]^{-1} ,
\label{H}
\end{equation}
with 
\begin{equation}
D\equiv\Rey^{-1}( \partial_{xx} + \partial_{yy} - \beta^2) - u_0\partial_x
- v_0\partial_y \, ,
\end{equation}
and the mean flow and corresponding spatial derivatives must be
provided for its construction. The resolvent operator acts as a
transfer function from nonlinearity $\bm{f}_{\beta,\omega}$ to
velocity fluctuations $\bm{u}_{\beta,\omega}$ in Fourier space. The
nonlinear terms are hence considered as the forcing that drives the
fluctuations. The gain properties of the resolvent are inspected via a
singular value decomposition (SVD). The resolvent operator is
factorized as
\begin{equation}
\mathcal{H}_{\beta,\omega}=\sum_m {\bm{\psi}}_{\beta,\omega,m}
\sigma_{\beta,\omega,m} {\bm{\phi}}^*_{\beta,\omega,m} \, ,
\label{eq:SVD}
\end{equation}
with $\bm{\psi}_{\beta,\omega,m}$ and $\bm{\phi}_{\beta,\omega,m}$
representing optimal sets of orthonormal singular response and forcing
modes respectively. These modes are ranked by the forcing-to-response
gain, {under the $L_2$ (energy) norm}, given by the corresponding
singular value $\sigma_{\beta,\omega,m}$. Here, the superscript $*$
indicates conjugate transpose and the subscript $m$ denotes the
ordering of the modes, from highest to lowest amplification.

The nonlinear forcing $\bm{f}_{\beta,\omega}$ can be projected onto
the orthonormal basis $\bm{\phi}_{\beta,\omega,m}$ to yield
\begin{equation}
\bm{f}_{\beta,\omega}= \sum_m
\bm{\phi}_{\beta,\omega,m}\chi_{\beta,\omega,m}
\end{equation}
where the unknown scalar coefficients $\chi_{\beta,\omega,m}$
represent the forcing sustaining the velocity fluctuations. The
introduction of this linear combination in conjunction with the SVD
(\ref{eq:SVD}) into the fluctuating velocity equation (\ref{eq:reseq})
reads
\begin{equation}
\bm{u}_{\beta,\omega} = \sum_m {\bm{\psi}}_{\beta,\omega,m}
\sigma_{\beta,\omega,m}\chi_{\beta,\omega,m}\, ,
\label{eq:reseq2}
\end{equation}
thus the velocity fluctuations in $(\beta,\omega)$ can be represented
as a linear combination of singular response modes weighted by an
unknown amplified forcing. Note that no assumption other than a
statistically steady flow has been employed in the derivation of
(\ref{eq:reseq2}), which is an exact representation of the
\NavSto\ equations.

\subsection{Rank-1 model reduction}
\label{sec.rank1}

A model reduction of the fluctuating velocity (\ref{eq:reseq2}) can be
carried out by considering the values taken by the amplification
$\sigma_{\beta,\omega,m}$. An inspection of these amplification
reveals that the first singular value $\sigma_{\beta,\omega,1}$ is
usually much larger than the second one $\sigma_{\beta,\omega,2}$,
hence the low-rank nature of the resolvent operator can be exploited
to yield a rank-1 model
\begin{equation}
\bm{u}_{\beta,\omega} \simeq {\bm{\psi}}_{\beta,\omega,1} a_{\beta,\omega,1}  \, ,
\label{eq:rank1}
\end{equation}
in which the product of amplification and forcing is collapsed into an
unknown complex amplitude coefficient
$a_{\beta,\omega,1}=\sigma_{\beta,\omega,1}\chi_{\beta,\omega,1}$. This
rank-1 model has proven to be adequate in previous investigations on
pipe and channel flows
\citep{McKeonSharma2010,SharmaMcKeon2013,Moarref2013,luhar2014opposition,gomez_pof_2014}.
Under this rank-1 assumption, the total velocity can be written as
\begin{equation}
\bm{u}(x,y,z,t) \simeq \sum_{\beta,\omega} a_{\beta,\omega,1} {\bm
  \psi}_{\beta,\omega,1} \ce^{\ci(\omega t - \beta z)} + \CC \,
\label{eq:rom} 
\end{equation}
We emphasize that this assumption is not required; any number of
singular response modes can be considered. However, this assumption
provides a convenient model in which the velocity fluctuations are
parallel to the first singular response mode. In what follows, we
rename the singular response modes as resolvent modes.

\subsection{Obtaining the amplitude coefficients}
\label{sec.coeffs}

The amplitude coefficients $a_{\beta,\omega,m}$ are unknown in the
model as a consequence of the closure problem. The present approach
can circumvent this issue by employing minimal additional information,
derived \eg from DNS.  Let us assume that the flow presents a number
$N_{\omega}$ of relevant frequencies $\omega_i$.  Under the rank-1
approximation, it follows from (\ref{eq:rom}) that the flow can be
reconstructed as a linear combination of resolvent modes
\begin{equation}
\sum_{i=1}^{N_{\omega}} {\bm{\psi}}_{\beta,\omega_i,1}(\bm{x})
a_{\beta,\omega_i,1} \ce^{\ci \omega_i t} = \bm{u}_{\beta}(\bm{x}, t)
\, ,
\label{eq:ls1}
\end{equation}
at any spatial location, time $t$ and wavenumber $\beta$. Owing to the
use of Fourier expansions in the spanwise direction, the closure
problem can be decoupled for each spanwise wavenumber $\beta$. The
linear system (\ref{eq:ls1}) contains $N_{\omega}$ unknowns and
consists of $3(N_xN_y)^2N_t$ scalar equations, with $N_t$ being the
number of time snapshots considered, hence their solution is amenable
to a least-squares approximation. While the terms in (\ref{eq:ls1})
are complex, it may be decoupled into real and imaginary equations. We
note that $3(N_xN_y)^2N_t \gg N_{\omega}$, hence the problem can be
restricted to a single spatial location $\bm{x}_0$ at a few instants
$N_t$
\begin{equation}
\sum_{i=1}^{N_{\omega}} {\bm{\psi}}_{\beta,\omega_i,1}(\bm{x}_0) a_{\beta,\omega_i,1} 
\ce^{\ci \omega_i t} = \bm{u}_{\beta}(\bm{x}_0, t) \, .
\label{eq:ls2}
\end{equation}
This has the advantage that the size of the problem is greatly reduced
and the temporal complexity is better captured. An interesting analogy
to experiments that will be introduced in \S\,\ref{sec.rom} is that
only a single velocity probe is required to obtain the unknown
amplitude coefficients. The problem could be further reduced by
considering only one velocity component. The least-squares solution of
(\ref{eq:ls2}) in matrix form is given by
\begin{equation}
\mathcal{A}_{\beta} = {\bm \Psi}_{\beta}^+{\mathcal{U}}_{\beta}(\bm{x}_0, t) \, 
\label{eq:ls4}
\end{equation}
with the $3N_t\times N_\omega $ matrix ${\bm \Psi}_{\beta}$ containing
the values of the resolvent modes at the spatial location $\bm{x}_0$
and different times, the $N_\omega \times 1$ vector
$\mathcal{A}_{\beta} $ represents the unknown amplitude coefficients,
and the $3N_t \times 1$ vector ${\mathcal{U}}_{\beta} $ contains the
values of the velocity at the spatial location $\bm{x}_0$ and
different times. The superscript $+$ denotes pseudo-inverse. As we
will show next, the typical dimensions of the least-squares problem
(\ref{eq:ls4}) are small and their solution is straightforward.

The proposed method relies on the modes being non-negligible at the
probe positions, hence these locations should be chosen
carefully. However, this limitation is may be readily overcome by
choosing a few probe positions at the locations of maximum velocity
for each mode. This also means it is possible to trivially fix the
amplitude coefficients from experimental data, since the probe
locations are provided by the peaks in the resolvent modes. One
strength of the present method is that it can recover relative phases
between the resolvent modes.

In principle, the fitting of the amplitude coefficients could be
carried out in physical space, but in the present work we have
exploited the fact that the probe data were obtained from a DNS, hence
a representation of the velocity signal in Fourier space
$\bm{u}_{\beta}(\bm{x}_0, t)$ has been employed in (\ref{eq:ls2}) to
fit simultaneously real and imaginary parts of the resolvent
modes. This has the advantage that two independent equations are
obtained at each time step considered. A least-squares fitting carried
out in physical space would need to consider twice the number of time
steps to obtain the same number of equations.

A least-squares fitting in Fourier space would not be possible if the
probe data were to be experimentally obtained. In that case, the
fitting could be carried out directly in physical space using
(\ref{eq:rom}) with probe data in physical space $\bm{u}(\bm{x}_0,
t)$. As such, a number $N_{\beta}$ of relevant wavenumbers $\beta_j$
could be simultaneously taken into account to yield the linear system
of real equations
\begin{equation}
\sum_{j=1}^{N_{\beta}} \sum_{i=1}^{N_{\omega}}
    {\bm{\psi}}_{\beta_j,\omega_i,1}(\bm{x}_0) a_{\beta_j,\omega_i,1}
    \ce^{\ci(\omega_i t - \beta_j z_0)} + \CC = \bm{u}(\bm{x}_0, t)
    \, ,
\label{eq:ls5}
\end{equation}
that could be solved in a least-squares sense as in (\ref{eq:ls4}).

\section{Reduced-order model of the cavity flow}
\label{sec.rom}

As outlined in \S\,\ref{sec.flow} the fluctuating velocity is
generated by three pairs of TGL vortices with different frequencies,
hence it is reasonable to focus on $\beta=3$. Note that although
figure \ref{fig:mdl} indicates non-zero fluctuating velocity at
$\beta=0$ and $\beta=6$, we will only consider $\beta=3$. The
$\omega=0$ contribution to the mode $\beta=0$ is the mean flow and it
is taken into account as an input in the resolvent. On the other hand,
the spanwise velocity fluctuations in $\beta=0$ are a consequence of
employing a spanwise periodic domain and lack physical meaning as they
could have been suppressed by imposing reflection symmetry at
$\beta=0$. The presence of end-walls would also prevent the existence
of these motions. The resolvent operator (\ref{H}) shows that the $w$
component can be decoupled from the rest of components at $\beta=0$,
hence the flow admits non-trivial solutions of $( D + \ci \omega ) w =
0$ as oscillations with an infinite span. The fluctuating velocity at
$\beta=0$ is dominated by $w$, hence the Reynolds stress contribution
to the mean flow $(u_0,v_0)$ is negligible. Additionally, figure
\ref{fig:mdl} indicates that the energy contained in $\beta=6$ is two
order of magnitude smaller than the one at $\beta=3$, hence a model
based on $\beta=3$ can provide a good representation of the
fluctuating spanwise velocity.

\begin{figure}
\begin{center} 
\includegraphics[width=0.725\linewidth]{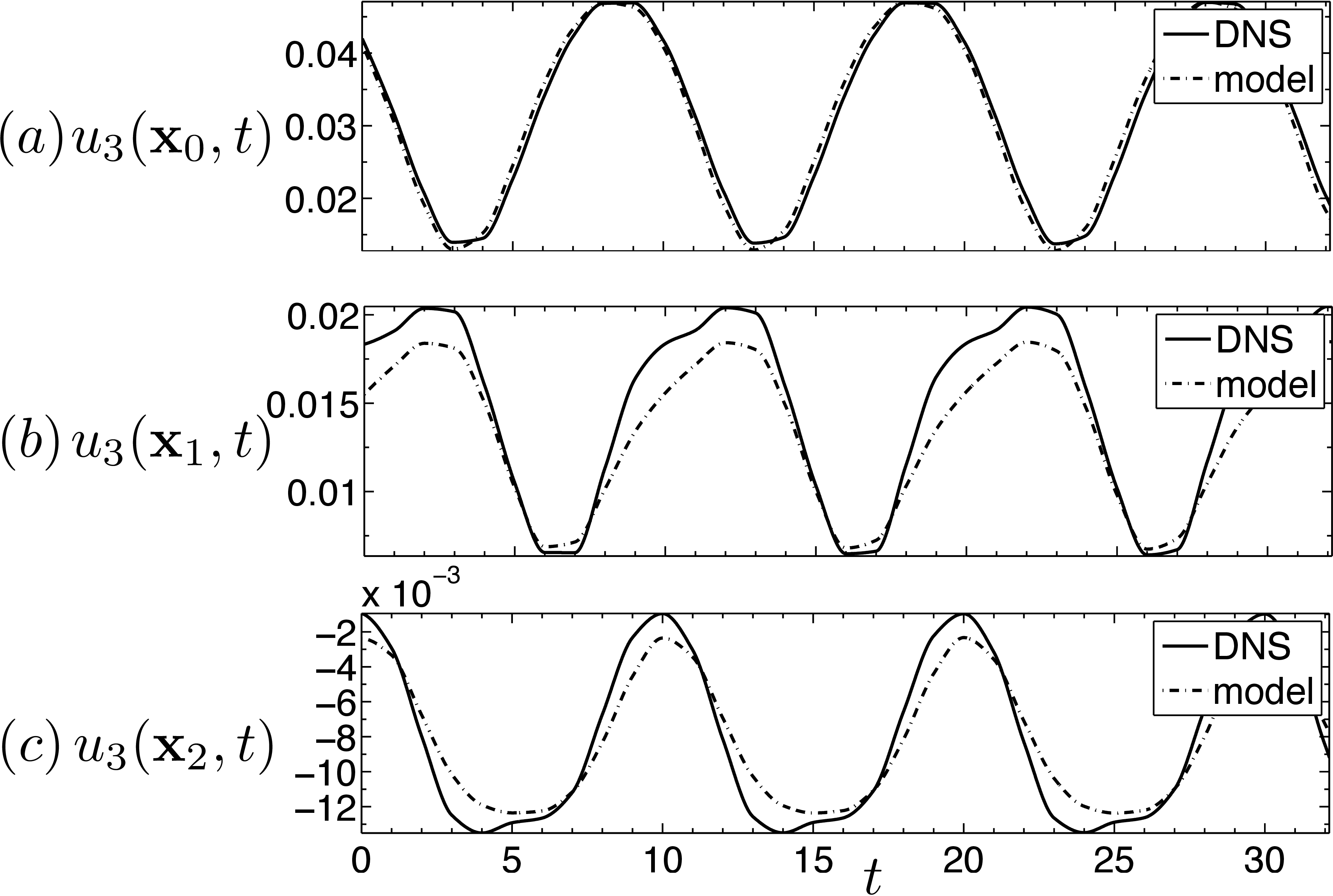} 
\end{center}
\caption{({\em a}) Temporal evolution of the real part of the velocity
  component $u$ at the location ${\bm x}_0=(0.1,0.1,0)$ at $\beta=3$
  obtained via DNS. The dashed line indicate the temporal evolution of
  the resolvent-based model at ${\bm x}_0$ with probe location at same
  point.({\em b}) DNS and model signals at ${\bm x}_1=(0.82,0.95)$
  with probe location at ${\bm x}_0$ ({\em c}) DNS and model signals
  at ${\bm x}_2=(0.4,0.26)$ with probe location at ${\bm x}_0$.}
\label{fig:frequencies}
\end{figure}

The problem may be decoupled for each $\beta$ as shown by
(\ref{eq:2.5D}). The first step in the construction of the model
consists of identifying the active frequencies in the flow. Figure
\ref{fig:frequencies}(a) presents the temporal evolution of the
velocity component $u$ at the location ${\bm x}_0=(0.1,0.1,0)$ at
$\beta=3$ obtained from the DNS. A temporal Fourier transform of this
signal indicates three active frequencies $\omega_0 = 0$, $\omega_1 =
0.76$ and $\omega_2 = 1.52$. A SVD of the resolvent operator is
carried out at $\beta=3$ for each of these active frequencies in order
to obtain the corresponding $(\beta,\omega)$ resolvent modes. The mean
flow at $Re=1200$ and $\Lambda/D=0.945$ obtained via DNS is employed
in forming the resolvent.

Note that the weightings of these spatial shapes are unknown. In order
to address this, the velocity values at ${\bm x}_0$ of these flow
structures are fitted to the velocity shown in
figure~\ref{fig:frequencies} via the least-squares problem
(\ref{eq:ls4}). This allows the amplitude of each of the resolvent
modes to be obtained and the construction of a reduced-order model
following (\ref{eq:rom}). As was anticipated, the size of the
least-squares problem (\ref{eq:ls4}) is small. In the present case, 40
measures of {the three velocity components} in a probe are employed to
fit three frequencies, hence the size of the matrix ${\bm \Psi_\beta}$
in (\ref{eq:ls4}) is only $120\times 3$.

\begin{figure}
\begin{center} 
\includegraphics[width=0.625\linewidth]{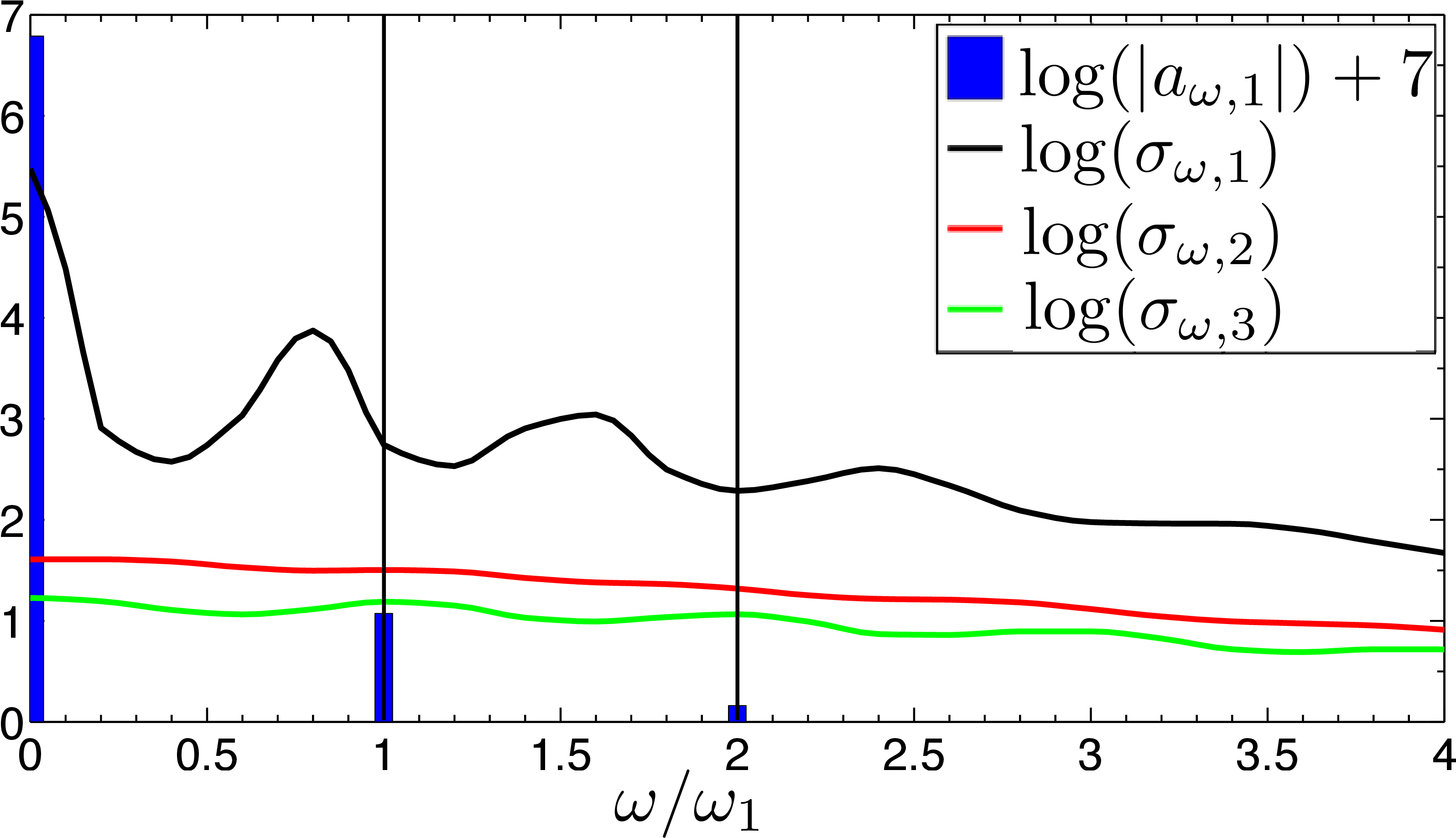} 
\end{center}
\caption{{Bars correspond to the (scaled) amplitude of the resolvent
    modes associated to each frequency at $\beta=3$. Lines denote the
    distribution in frequency of the first, second and third singular
    value at $\beta=3$.}}
\label{fig:sigma}
\end{figure}

The bars in figure \ref{fig:sigma} show the value of the amplitudes
$a_{\beta,\omega,1}$ of each of the three modes considered at
$\beta=3$ obtained via the fitting. A large decrease of the amplitude
is observed with increasing frequency, justifying omission of higher
harmonics. In addition, figure~\ref{fig:sigma} shows the distribution
of the first, second and third singular value in frequency.  The
rank-1 assumption is justified by the fact that the first singular
value is always orders of magnitude larger than the rest.  We note
that peaks in amplification do not necessarily correspond to peaks in
amplitude or, in this case, even to active frequencies. This has been
previous observed by \citet{moarref2014low} and \citet{gomez_iti_2014}
in turbulent canonical flows. A connection with these amplification
peaks and the concept of optimal forcing can be found in the work of
\citet{monokrousos2010global}.

\begin{figure}
\begin{center} 
\includegraphics[width=0.9\linewidth]{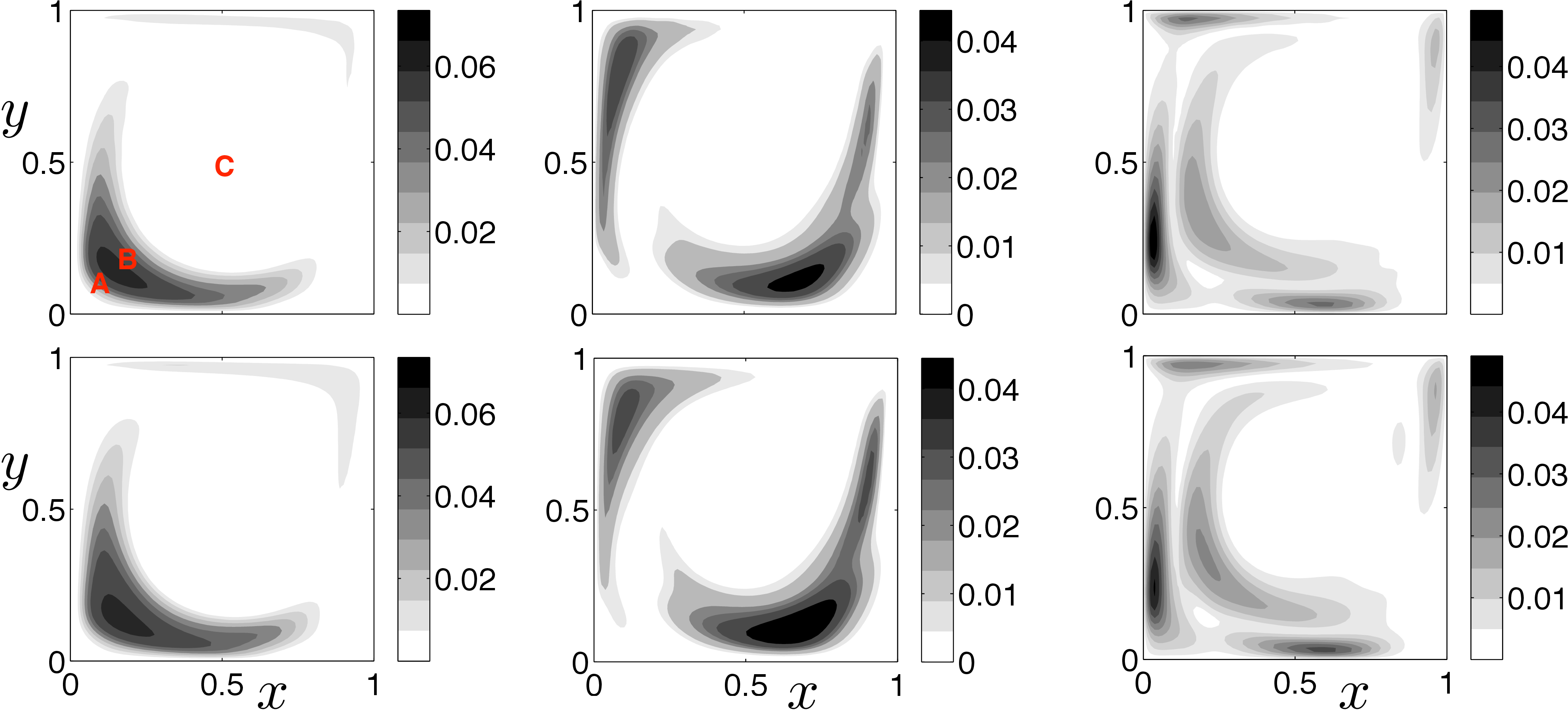} 
\end{center}
\caption{Comparison of the fluctuating velocity intensities (root-mean
  square) obtained from (upper) DNS (lower) Resolvent-based
  model. Left to right: $u_{rms}$, $v_{rms}$ and $w_{rms}$. }
\label{fig:intensities}
\end{figure}

Figure \ref{fig:intensities} shows a comparison between the
fluctuating intensities obtained from the DNS and the present
reduced-order resolvent-based model.  It is observed that the regions
of maximum fluctuating intensity predicted by the resolvent model are
in good agreement with those observed by DNS. Also, the values of the
maximum fluctuation intensities could be exactly recovered if the
probe is located at the corresponding spatial location of the maxima.
Additionally, we observe that the maxima of fluctuating intensities
are close to $7$\% of the lid speed, hence these are not negligible
with respect to the mean flow.

Figure \ref{fig:frequencies} shows the temporal evolution of the
resolvent-based model at different locations with a probe fixed at
$\bm{x}_0$ and a comparison with the corresponding DNS signals. These
location correspond to regions in figure \ref{fig:intensities} in
which the $x$-velocity component $u$ is significant. We observe that
the model accurately recovers the DNS signal at the probe location, as
it was expected. Although small discrepancies in the two other
locations are visible, the shape and phase of the DNS signals are in
good agreement.

Figure \ref{fig:LDC}(\textit{b}) shows a reconstruction of the flow
using the present resolvent-based model. A good agreement between the
model and the Fourier mode corresponding to $\beta=3$ obtained from
DNS is observed. We observe that the flow structures corresponding to
the TGL vortices are well recovered. In order to quantitatively
measure the error of the model with respect to the DNS, we define a
percentage error as
\begin{equation}
\Delta\bm{u}(\%) = 100\times \|\bm{u}_\beta - \bm{u}_\beta^R \|/ 
\|\bm{u}_\beta\| \,
\label{eq:error}
\end{equation}
where $\|.\|$ denotes $L_2$ norm and $\bm{u}_\beta^R$ is the
resolvent-based model arising from the least-squares approximation
(\ref{eq:ls4}). Table~\ref{tab:error} presents numerical values of the
error using different spatial locations for the probe.

\begin{table}
\centering 
\begin{tabular}{c c c | c c c c }
Probe & $\bm{x}_0$ &  $\Delta\bm{u}$ (\%) & $\Delta u$  (\%) & $\Delta v$  (\%) & $\Delta w$ (\%) \\ 
$A$ \; &  (0.1,0.1,0)  \;  & 3.77\; & 3.19\; &  6.48\;  & 5.03  \\ 
$B$ \; &  (0.2,0.2,0)  \;  & 3.97\; & 3.19\; &  6.77\;  & 5.74  \\ 
$C$ \; &  (0.5,0.5,0)  \;  & 870.25\; & 1138.03 \; &  1019.57 \;  & 830.16  \\
$A+C$ \; &  -  \;  & 3.77\; & 3.18\; &  6.47\;  & 5.04 \\ 
\end{tabular}
 \caption{\label{tab:error} Measure of error of the model based on
   (\ref{eq:error}). Probe positions are shown in
   figure~\ref{fig:intensities}.}
\end{table}

As mentioned in \S\,\ref{sec.model}, the resolvent modes provide the
spatial locations in which the fluctuating velocity is expected to be
significant. We observe a consistent error of around 4\% by choosing
probe locations based on this information, like the probes $A$ and
$B$.  On the other hand, if the probe is placed in a location in which
the fluctuating velocity is small, such as probe $C$ in the center of
the cross section cavity, the model can fail. However, this limitation
can be overcome by employing more than one probe. For instance, a
combination of probes $A+C$ yields an error of around 4\%. As the
number of probes employed increases, the fitting results can be
improved. This leads us to believe that the best results could be
obtained if full velocity snapshots are employed. We highlight that
DMD of statistically steady data provides a representation of the flow
equivalent to the Fourier transform (\ref{eq:2.5D}), hence the present
resolvent-based model would approach DMD as the number of probes and
the rank of the model increases. However, if complete velocity
snapshots are available, an empirical analysis such as DMD would be
the best tool to obtain a representation of the the flow, since the
model can be directly extracted from post-processing of the available
data.

On the other hand, if only the mean flow and local (one probe)
information are available, the present method could be the tool of
choice to construct a ROM. An example of this scenario can be found in
experiments dealing with canonical geometries, such as pipe or channel
flows. The mean flow is typically measured using a hot wire anemometer
at different wall-normal distances in order to obtain statistics of
the entire profile. As such, snapshots of the flow are not obtained
but time-histories of the velocity at selected locations are
available. Another scenario in which only a mean flow and probe
information are available would be a combination of simulations and
experiments. An approximation to the mean flow could be computed via
RANS, as in \citet{meliga2012sensitivity}, while the spectral
information could be experimentally obtained from pressure or velocity
measurements.

In terms of model reduction, the DNS consists of 32 Fourier modes in
the spanwise direction (64 planes), while the resolvent-based model
has three modes (6 planes). A similar model reduction would be
obtained from Fourier analysis or DMD applied to the DNS
dataset. However, the DNS requires many planes because the nonlinear
terms are evaluated in physical space, hence a good resolution in the
spanwise direction is required, even if most of those 32 modes have
zero amplitude.

The present representation of the fluctuating velocity in conjunction
with the mean flow equation~(\ref{mean}) represents a low-dimensional
dynamical system susceptible to flow control studies. This flow
control framework has been successfully employed by
\citet{luhar2014opposition,luhar2015framework} to investigate
opposition flow control and the effect of compliant walls in a
turbulent pipe flow by incorporating additional forcing in the
resolvent equation~(\ref{eq:reseq}). Despite the assumptions employed
in those works, namely (i) mean-flow not affected by control, (ii)
unit-broadband forcing, (iii) rank-1 model truncation and (iv)
statistically-steady flow, the flow control results were
satisfactory. We note that the current approach makes unnecessary the
assumption (ii); the present model provides amplitudes to the
resolvent modes. Removing the assumptions (i) and (iv) remains a
future challenge.

\section{Conclusions}
\label{sec.conc}

A reduced-order model of an unsteady lid-driven cavity flow based on
the resolvent decomposition has been developed. It requires only the
mean flow and minimal spectral information to help identify relevant
frequencies.  In the present application, where the flow had a
homogeneous direction, temporal information from a single (but perhaps
re-positionable) probe was employed, but the most active spanwise
wavenumber was known in advance; in practice the identification of
this wavenumber would require simultaneous measurements from at least
two traversable probes.  Only a single probe would be required to
reconstruct a \threed\ flow without a homogeneous direction.

We have demonstrated that the model could predict the regions and
values of the fluctuating velocity intensities with an error of order
5\%. In addition, the model may be improved by employing additional
probes; the rank-1 approximation has proven to be useful, but one may
expect the error to reduce as the number of resolvent modes is
increased.
As opposed to global stability analysis, no assumptions concerning the
nonlinear terms are involved in the derivation of the model. Hence,
resolvent analysis seems most appropriate for constructing ROM of
flows which are neither weakly nonlinear nor fully turbulent.
The resolvent modes have an orthonormal basis, hence the
resulting ROM is not affected by non-normality of the operator.
The size of the least-squares problem to be solved in order to
construct the ROM is related to the number of active frequencies in
the flow. A low-order representation of broad-spectrum flows is left
for future work.

While in the present work DNS was employed both to obtain the mean
flow on which the resolvent analysis was based and to provide
comparison data, we emphasize that in principle these two steps could
be separated: the mean flow and spectral information could be obtained
independently.

\section*{Acknowledgments}
The authors acknowledge financial support from the Australian Research
Council through grant DP130103103, and from Australia's National
Computational Infrastructure via Merit Allocation Scheme grant D77.

\bibliography{cavity.bib}
\bibliographystyle{jfm}

\end{document}